\documentclass[aps,prx,reprint,amsmath,amssymb]{revtex4-2}

\usepackage[bookmarks=false,linkcolor=NavyBlue,urlcolor=NavyBlue,colorlinks,citecolor=NavyBlue]{hyperref}
\usepackage{graphicx}
\usepackage{amsmath}
\usepackage{bm}
\usepackage[svgnames]{xcolor}
\usepackage{enumerate}
\usepackage[normalem]{ulem}
\usepackage{microtype}
\usepackage{comment}

\renewcommand{\vec}[1]{\mathbf{#1}}

\newcommand{\figref}[2]{\hyperref[#1]{\ref{#1}(#2)}}
\newcommand{\figsref}[2]{\hyperref[#1]{\ref{#1}#2}}

\graphicspath{{./figures/}}

\begin{document}

\newcommand{\Cornell}{Department of Physics, Cornell University, Ithaca, NY 14853, USA}

\newcommand{\EqualContribution}{These authors contributed equally to this work.}

\title{Charting the emergent low-dimensional manifold of quantum materials}

\author{Jason Z. Kim}\thanks{\EqualContribution}
\author{Omri Lesser}\thanks{\EqualContribution}
\author{Debanjan Chowdhury}\email{debanjanchowdhury@cornell.edu}
\affiliation{\Cornell}

\begin{abstract}
The periodic table of elements transformed chemistry by revealing simple organizing principles underlying atomic behavior. Despite decades of effort, no analogous framework has emerged for crystalline materials --- their microscopic complexity and vast configurational space have defied reduction to fundamental organizing principles. Current databases catalog thousands of synthesized materials, but extracting predictive, interpretable models from this wealth of data remains a formidable challenge. Here we demonstrate that the materials landscape possesses a hidden geometric organization that can be unveiled through {\it unsupervised} nonlinear dimensionality reduction. Applying differential geometry techniques to the Inorganic Crystal Structure Database (ICSD), we reveal that just a few combinations of microscopic descriptors capture the vast majority of variance in material properties. This low-dimensional embedding autonomously segregates superconductors from ordinary materials and further distinguishes superconducting families in ways that transcend chemical similarity alone. Remarkably, the discovered geometric organization directly governs critical temperatures ($T_c$) across diverse superconducting families, enabling accurate $T_c$ predictions without any knowledge of the pairing mechanism. Our approach uncovers emergent organizing principles that control macroscopic quantum behavior, offering a new paradigm in how we build models of complex quantum materials directly from experimental data.
\end{abstract}
\maketitle
\tableofcontents
\section{Introduction}
From mapping amino acid sequences to protein structure~\cite{jumper_highly_2021,tunyasuvunakool_highly_2021}, to navigating the landscape of human language and reasoning~\cite{noy_experimental_2023,vaswani_attention_2017,openai_gpt-4_2024}, some of the most profound advances in modern times involve distilling emergent function from complex data through models. The corresponding models then serve as a map for exploring the structure of emergence from the quantitative descriptors in the data~\cite{meng_locating_2022,cunningham_sparse_2023,rao_transformer_2020,zhang_protein_2024}. An analogous and central challenge in electronic structure theory has been inferring physical properties of complex materials from their chemical composition and structure. While the periodic table helps organize chemical understanding of elements, predicting physical observables for specific compounds from first principles remains challenging. For instance, distinguishing metals from insulators requires electronic structure calculations, which are computationally intensive and often unreliable for strongly interacting systems. Moreover, finding patterns across materials typically requires repeating these calculations entirely. This complexity stems from the vast combinatorial space of electronic parameters (e.g., electronegativity, ionization energies) and structural parameters (e.g., bond lengths and angles) in multi-element compounds.

We conjecture that only a few microscopic features determine the geometric organization of experimentally stable materials within this combinatorial space. By extending differential geometry principles to high-dimensional spaces, we verify this conjecture and demonstrate its utility by analyzing how the superconducting materials are organized relative to the rest in the Inorganic Crystal Structure Database (ICSD)~\cite{zagorac_recent_2019}. Predicting superconductivity and the associated critical temperature ($T_c$) from first principles remains a fundamental challenge, largely because we lack a unified understanding --- even statistically --- of which chemical and structural properties govern superconducting behavior. This gap in our understanding makes the problem an ideal testbed for our approach.

Our method builds on two key resources: the curated ICSD database, which contains over 220{,}000 known stable compounds~\cite{zagorac_recent_2019}, along with their associated properties~\cite{sommer_3dsc_2023,chen2024learning}; and a systematic framework for embedding these compounds into a unified, high-dimensional feature space derived from atomic properties and crystal structure~\cite{merchant_scaling_2023,gashmard2024predicting,chen2024learning,cheng_ai-driven_2025,cheng_enhancing_2025,wang_experimentally_2025,ma_topogivity_2023,ma_learning_2025,lesser_learning_2025,panigrahi_graphlet_2026}. By sampling this broad chemical space, we extract emergent patterns in the materials landscape directly from empirical data.

The central challenge is discovering a low-dimensional organization of the materials landscape that is both predictive and interpretable --- ideally in terms of a few physically meaningful coordinates. While supervised machine learning effectively predicts $T_c$ from microscopic features~\cite{stanev_machine_2018,hamidieh_data-driven_2018,roter_predicting_2020,pereti_individual_2023,pogue_closed-loop_2023,gashmard2024predicting,chen2024learning,kaplan_deep_2025,xie_machine_2022,gibson_developing_2025,prakash_guided_2025}, the resulting nonlinear mappings obscure which structural or chemical principles contribute most to superconductivity. Recent work has identified specific interpretable features that correlate with $T_c$ across superconducting families~\cite{lesser_learning_2025}. However, there is clear empirical evidence that suggests multiple sparse feature subsets achieve comparable accuracy~\cite{lesser_learning_2025}, implying that the materials landscape admits organization along several {\it collective axes}~\cite{tahmassebpur_seven-facet_2026}.

Standard unsupervised dimensionality reduction techniques like UMAP~\cite{mcinnes2018umap,ye2022novel} map high-dimensional data to low-dimensional latent spaces but produce embeddings that distort distances and lack interpretable structure. Similarly, autoencoders~\cite{lee2022powder} learn continuous manifolds but remain uninterpretable; small shifts in latent coordinates can produce qualitatively different predictions of material features~\cite{arvanitidis2017latent,kim2024gamma}, undermining both interpretability and generalization. 

We therefore need unsupervised methods that construct low-dimensional models with their geometric meaning preserved, such that the latent axes correspond to regular variations in the collective features. Such models would enable a systematic and principled exploration of material properties. If materials in the latent space organize along an increase in a property such as $T_c$, that organization would have resulted from systematic and regular changes in the collective features that we could directly decipher from our model. These collective features could then serve as putative leading-order terms in developing a broader empirical theory of $T_c$. To achieve the geometric regularity necessary for such models, we must penalize irregularities and distortions of the manifold.

To address these challenges, we use the $\Gamma$-autoencoder ($\Gamma$AE)~\cite{kim2024gamma} --- a data-driven technique using differential geometry to control distortions causing overfitting --- to unveil a novel geometric organization for materials. The resulting low-dimensional manifold smoothly follows the structure of the data in high-dimensional feature space, with materials projections onto manifold coordinates revealing organization along interpretable collective axes. Remarkably, this geometric structure separates most superconductors from other materials and reveals microscopic feature combinations correlating with higher $T_c$ within superconducting families. When trained without superconductors, the geometric embedding of superconductors relative to remaining materials stays unchanged, demonstrating autonomous learning of the patterns leading to superconductivity. The interpretable embedding enables analyzing features importance for $T_c$, thereby providing a complementary discovery approach free from various underlying theoretical approximations.

The remainder of this manuscript is organized as follows.
In Sec.~\ref{subsec:data} we present the materials database and the method for generating the descriptors. In Sec.~\ref{subsec:gammaae} we describe our geometry-aware autoencoder architecture. We then present the resulting embedding of materials in Sec.~\ref{subsec:embedding} and discuss the emergent clustering of materials. In Sec.~\ref{subsec:geometric_sc} we further develop a geometric approach to superconductivity based on this embedding. We conclude and discuss possible implications in Sec.~\ref{sec:discussion}.

\section{Preliminaries}

\subsection{Materials: data and features}\label{subsec:data}

\begin{figure*}
    \centering
    \includegraphics[width=\linewidth]{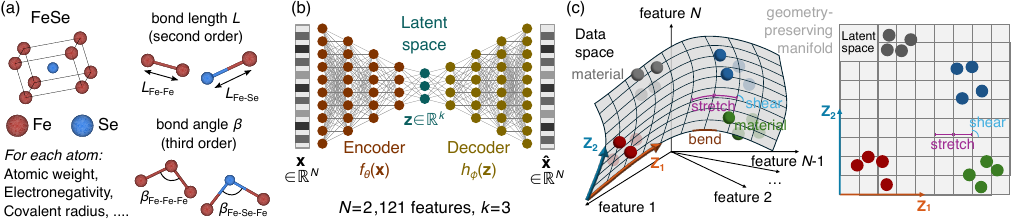}
    \caption{ {\bf Featurization, $\boldsymbol{\Gamma}$AE, and geometry preserving manifolds.}
    (a)~Up to third-order graphs are used to include features that are products of atomic properties (see Table~\ref{tab:symbols}) of each node.
    (b)~Data ($\bm{x}$) is compressed to $\bm{z}$ in the latent space by the encoder neural network, followed by a decoder neural network to minimize the difference between the reconstructed data ($\hat{\bm{x}}$) and $\bm{x}$, while regulating the geometry (see Appendix~\ref{sec:gvae}).
    (c)~The $\Gamma$-autoencoder ($\Gamma$AE) constructs a geometry-preserving, low-dimensional manifold in feature space (left), using the latent space (right) as its coordinate system, so that distances and angles are preserved and the manifold bends smoothly along latent-space directions.
    }
    \label{fig:intro_method}
\end{figure*}

We use the atomic and structural information of each compound to construct features based on a graph expansion of the crystal structure~\cite{scarselli_graph_2009,isayev_universal_2017,bronstein_geometric_2021,reiser_graph_2022,fung_benchmarking_2021,zhang2023gpt}. These are hierarchical sets of features, starting from the properties of individual atoms and building up clusters of an increasing number of atoms (see Fig.~\figref{fig:intro_method}{a} for an illustration for a specific compound)~\cite{isayev_universal_2017,choudhary_atomistic_2021,lesser_learning_2025,panigrahi_graphlet_2026}, providing a comprehensively detailed but simplified numerical description of the material. A few representative examples of atomic features include the atomic weight, electronegativity, covalent radius, ionization potential, etc., while geometric properties include bond lengths ($L$) and bond angles ($\beta$); see Table~\ref{tab:symbols} for a full list. 

Unless otherwise stated, our results will be based on a third-order (three-atom cluster) featurization, where for each cluster we include all combinations of products of atomic and geometric properties and use the first five moments combined with the minimal and maximal values as our features. This yields a relatively large $2{,}121-$dimensional feature vector for each material (see Appendix~\ref{sec:sm_featurization}). We note that while distinct featurizations can modify the structure of the emergent geometric organization, the results remain qualitatively similar (see Appendix~\ref{sec:different_featurizations}).

\begin{table}
\centering
\begin{tabular}{cc}
\hline
\textbf{Symbol} & \textbf{Meaning} \\
\hline
$M$ & Atomic mass \\
$R_{\rm A}$ & Atomic radius \\
$R_{\rm C}$ & Covalent radius \\
$\chi$ & Pauling electronegativity \\
$A$ & Electron affinity \\
$E_{I}$ & First ionization energy \\
$N_{\rm v}$ & Number of valence electrons \\
$G$ & Group in the periodic table \\
$P$ & Period in the periodic table \\
$L$ & Bond length \\
$\beta$ & Bond angle \\
\hline
\end{tabular}
\caption{List of atomic and geometric properties used for $\Gamma$AE featurization in this study.}
\label{tab:symbols}
\end{table}

\subsection{$\boldsymbol{\Gamma}$-Autoencoder architecture}\label{subsec:gammaae}
We use a deep neural network, \textit{$\Gamma$-autoencoder}, that condenses the data onto a low-dimensional geometric manifold while penalizing sharp distortions in the manifold geometry [Fig.~\figref{fig:intro_method}{b,c}]. Specifically, we build our model manifold using an autoencoder whose architecture comprises an encoder, $\bm{z} = f_\theta(\bm{x})$, which maps material samples from $N$-dimensional feature space to a $k$-dimensional latent space, $\mathbb{R}^N \rightarrow \mathbb{R}^k$, where $k\ll N$, and a decoder, $\hat{\bm{x}} = h_{\phi}(\bm{z})$, which maps latent space samples back to feature space, $\mathbb{R}^k \rightarrow \mathbb{R}^N$ (see Fig.~\figref{fig:intro_method}{b} for a schematic~\cite{kingma2019introduction,li2023comprehensive}). Here, $f_\theta(\cdot)$ and $h_\phi(\cdot)$ are four-layer perceptrons using softplus activation functions. Unless noted otherwise, we will show results for $N=2{,}121$ and $k=3$ (see Appendix~\ref{sec:different_featurizations} for additional results).

The low-dimensional model manifold of predictions is parameterized by the latent space coordinates, and aims to capture the underlying low-dimensional structure that organizes the data. Since deep neural networks are universal function approximators~\cite{hornik1989multilayer}, the geometry of the manifold defined by typical neural networks is significantly distorted~\cite{arvanitidis2017latent}. Specifically, the map between the latent space coordinates and the model manifold of predictions of material features becomes distorted because the manifold is allowed to stretch and bend to visit the high-dimensional data in any arbitrary order~\cite{kim2024gamma}. As a result, interpreting the specific ordering of points in the latent space and the corresponding changes tied to the collective features becomes difficult. Most importantly, identifying the embedding for new materials outside of the training data becomes unreliable.

We extend the recently developed $\Gamma$AE~\cite{kim2024gamma} algorithm to the present setting, which regularizes the emergent geometry by performing a complete and orthogonal decomposition of all nonlinearities present in the manifold. This decomposition involves two curvatures: the \textit{parameter-effects curvature} that distorts the metric along the manifold tangents, and the \textit{extrinsic curvature} that allows the low-dimensional manifold to bend sharply in new directions in materials space (see Appendix~\ref{sec:gvae}). By controlling the parameter-effects curvature, a regular grid of points in the latent space maps to a regular grid of points on the manifold in data space [Fig.~\figref{fig:intro_method}{c}], which ensures that distances and angles at two points in the latent space do not map to dramatically different distances and angles on the manifold in data space. If the latter is not accounted for, then the directions in data space that are traversed can change rapidly as we move smoothly along the latent space, making it difficult to interpret the latent space axes with respect to the data features. By controlling the extrinsic curvature, we ensure that a straight line in the latent space corresponds to a gently bending curve in the data space, such that the axes of the latent space correspond to slowly-varying collective features [Fig.~\figref{fig:intro_method}{c}].

While the superconducting attribute and its $T_c$  are \emph{not} part of the training process, once we obtain the low-dimensional embedding, we use the database 3DSC-ICSD~\cite{sommer_3dsc_2023} to label the superconducting compounds. This is based on matching entries in ICSD to entries of the SuperCon database~\cite{noauthor_mdr_nodate}, which tabulates all known three-dimensional superconductors and their critical temperatures $T_{c}$.

\section{Results}
\subsection{Emergent three-dimensional embedding}\label{subsec:embedding}

\begin{figure*}
    \centering
    \includegraphics[width=\linewidth]{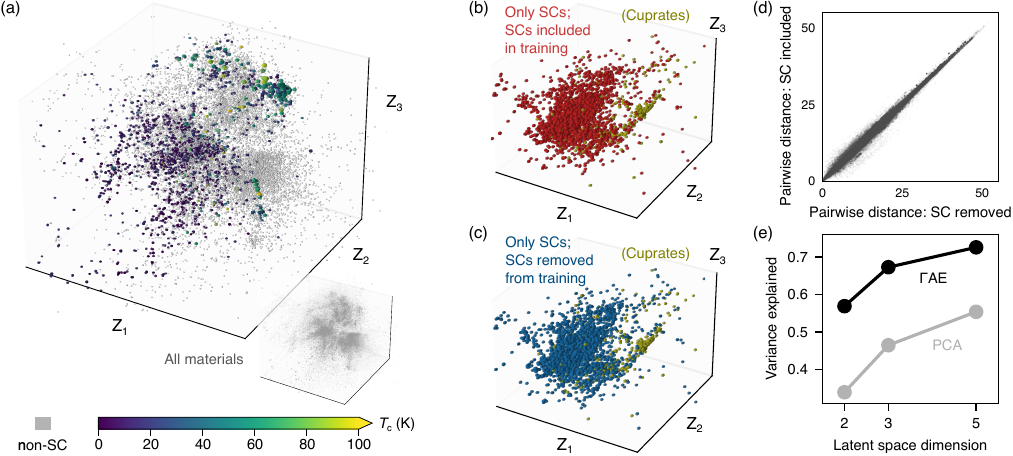}
    \caption{{\bf Three-dimensional embedding.} 
    (a)~The embedding for all ICSD materials features three main clusters (see inset), with a clear organization in terms of their superconducting $T_c$.
    (b--c)~Embedding of \emph{only} the superconducting materials when they are (b) included vs. (c) excluded from the $\Gamma$AE training process.
    The autoencoder consistently embeds the superconducting materials, including individual families (e.g. cuprates) in the same location without even encountering them during training.
    (d)~Pairwise distance between each superconductor and $400$ randomly selected materials in the two embeddings in (b) and (c) demonstrates their high correlation (Pearson $r=0.998$).
    (e)~Comparison of the variance captured using $\Gamma$AE vs. (linear) PCA for 2, 3, and 5 latent space dimensions, respectively.
    }
    \label{fig:re_embedding}
\end{figure*}

The three-dimensional embedding obtained using our network is depicted in Fig.~\figref{fig:re_embedding}{a}, with the latent space coordinates labeled by $\{Z_1,Z_2,Z_3\}$. For our particular featurization scheme, we find that the materials embedding is organized into three well-separated clusters [see the inset of Fig.~\figref{fig:re_embedding}{a}].

We use the decoder to obtain the relationship between the latent space coordinates $\{Z_j\}$ and variations in the original features; however, due to the nonlinearity of the transformations in the autoencoder, these relationships are position-dependent within the embedding. For example, in the left-most cluster in the inset of Fig.~\figref{fig:re_embedding}{a}, our analysis reveals that the following {\it third-order} features vary most significantly along the three latent space coordinates, respectively: $(E_{\rm{I}}\times G\times P)$; $(M\times R_{\rm A}\times R_{\rm A})$; $(A\times L\times M)$. Here $E_{\rm I}$ is the ionization potential, $G~(P)$ is the group (period) in the periodic table, $M$ is the atomic mass, $R_{\rm A}$ is the atomic radius, $A$ is electron affinity, and $L$ is bond length (see Table~\ref{tab:symbols}). For the specific cluster, we have found that the relative variations of the above features along the respective latent space axes are approximately an order of magnitude larger than any other set of features. Note that the specific set of features that vary along the axes can vary from one cluster to another. Moreover, the number of features that vary collectively generically depends on the specific cluster; however, we have noticed that within each cluster only a few features vary locally in a significant fashion.

Labeling all the materials in Fig.~\figref{fig:re_embedding}{a} according to their superconducting $T_c$ reveals a clearly increasing trend across the low-dimensional embedding. Moreover, we can compare the embeddings associated with the superconducting materials when they were: (i) included in the autoencoder training dataset [Fig.~\figref{fig:re_embedding}{b}], (ii) excluded from the autoencoder training dataset [Fig.~\figref{fig:re_embedding}{c}]. Remarkably, the two embeddings appear to be nearly identical in terms of their overall structure and the relative placement of specific superconducting families (e.g., the cuprates). We have computed the mean distance between each superconducting compound and a set of $400$ randomly selected materials to compare their relative location across the two embeddings [Fig.~\figref{fig:re_embedding}{d}], and find that they match to a large degree. The Pearson correlation coefficient~\cite{pearson_note_1997} evaluates to $r=0.998$, suggesting an almost perfect linear correlation between the two embeddings. In Appendix~\ref{sec:different_featurizations} we show very similar results when \emph{all} compounds containing Cu and O are excluded from the autoencoder training dataset.

\begin{figure*}
    \centering
    \includegraphics[width=\linewidth]{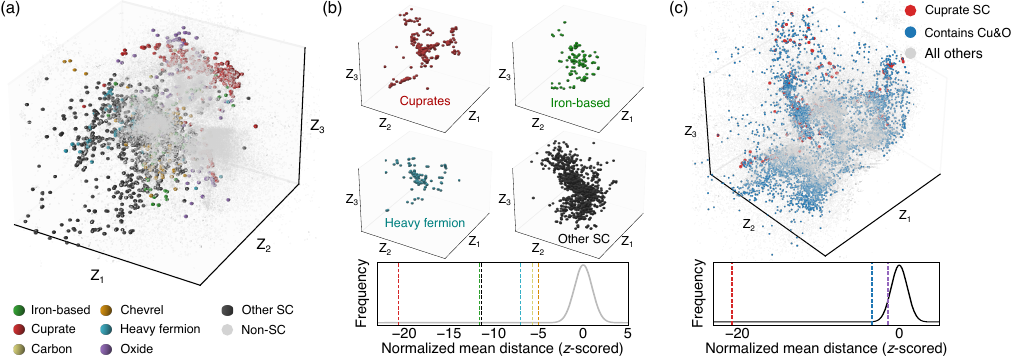}
    \caption{ {\bf Geometric clustering of superconducting families.}
    (a)~Superconducting materials colored by family across the embedding.
    (b)~The $z<-2$ scores for the pairwise distance, including for the ``other" (i.e. conventional) superconductors, demonstrates their significant clustering tendencies beyond chemical similarity.
    (c)~Clustering of superconducting cuprates (red dots) beyond compounds containing Cu and O (blue dots) 
    The in-group pairwise distance normalized by the group size is statistically significantly lower for superconducting cuprates compared to the latter. The purple dashed line is the $z$-score for cuprate superconductor distances relative to the distribution of distances of all compounds containing Cu and O.
    }
    \label{fig:3d-embedding}
\end{figure*}

From a data-driven perspective, one might rationalize these results by noting that removing the superconductors, which make up less than 2\% of the materials, from the training set should have no effect on the resulting embedding. On the other hand, the network is able to make out-of-distribution predictions, by distinguishing superconducting materials from the rest without ever having encountered the former during training. These results suggest that the $\Gamma$AE model can identify the handful of geometric and electronic properties that capture the organizational aspects of superconducting materials, and maximally distinguish them from the rest. We have also observed that with the current featurization, our embedding captures 67\% of the variance of the full features, compared to a mere 46\% captured by three-component principal component analysis (PCA)~\cite{jolliffe_principal_2016}; the explained variance in the data increases with increasing latent space dimension [Fig.~\figref{fig:re_embedding}{e}].

\subsection{Geometric theory of superconductivity}\label{subsec:geometric_sc}

Upon careful scrutiny of the embedding, we find that the superconducting cluster is further segregated into chemically distinct families [Fig.~\figref{fig:3d-embedding}{a}]. We illustrate this clustering by plotting only the cuprate, iron-based, heavy fermion, and ``other" (i.e. conventional) superconductors one at a time in Fig.~\figref{fig:3d-embedding}{b}. 

It is important to analyze the extent of clustering quantitatively and to examine if it is a natural consequence of solely their chemical similarity or superconductivity. For the first question, we have computed the mean pairwise distance between materials within a cluster using the standard Euclidean distance metric appropriate for our curvature-regulated embedding. This mean distance is then compared to the mean pairwise distance in a randomly generated ``cluster". For a presumptive cluster of $M$ materials, we selected a set of $M$ uncorrelated points as the null distribution (see Gaussian curve in bottom panel of Fig.~\figref{fig:3d-embedding}{b} and Appendix~\ref{sec:gradient}), with a variance that scales as $1/\sqrt{M}$. This process is repeated 10{,}000 times to generate the null distribution. Thus, to compare clusters of different size $M$, we standardize the distributions (via $z$-scoring the distances~\cite{freund_statistical_2003}), by subtracting the mean and dividing by the associated $M$-dependent variance.

We find that all superconducting families cluster significantly more than expected by chance ($z<-2$), which includes the ``other'' category [black dashed line in Fig.~\figref{fig:3d-embedding}{b}]. The last observation is especially remarkable, since they primarily include the conventional superconductors which are chemically dissimilar, and yet the unsupervised autoencoder clusters them together. To address the second question of the extent of clustering due to chemical similarity, we examine the specific example of cuprate superconductors. We find that they are significantly more clustered compared to all compounds containing copper (Cu) and oxygen (O); see Fig.~\figref{fig:3d-embedding}{c}. While compounds containing Cu and O show a score $z=-3.4$, the cuprate superconductors are clustered with a score $z=-20.7$. When compared to the distribution of distances within the Cu and O cluster, the cuprate superconductors have $z=-1.4$, which shows clustering but to a lesser degree.

The example of the conventional and cuprate superconductors demonstrates that our autoencoder embedding goes well beyond clustering materials solely based on their chemical formulae. The qualitative behavior persists when we choose a two-dimensional embedding and different featurization schemes (see Appendix~\ref{sec:2d_emb}), including the clustering of different superconducting families beyond chemical similarity and the overall organization of all materials into three main clusters.

\begin{figure*}
\centering
\includegraphics[width=\linewidth]{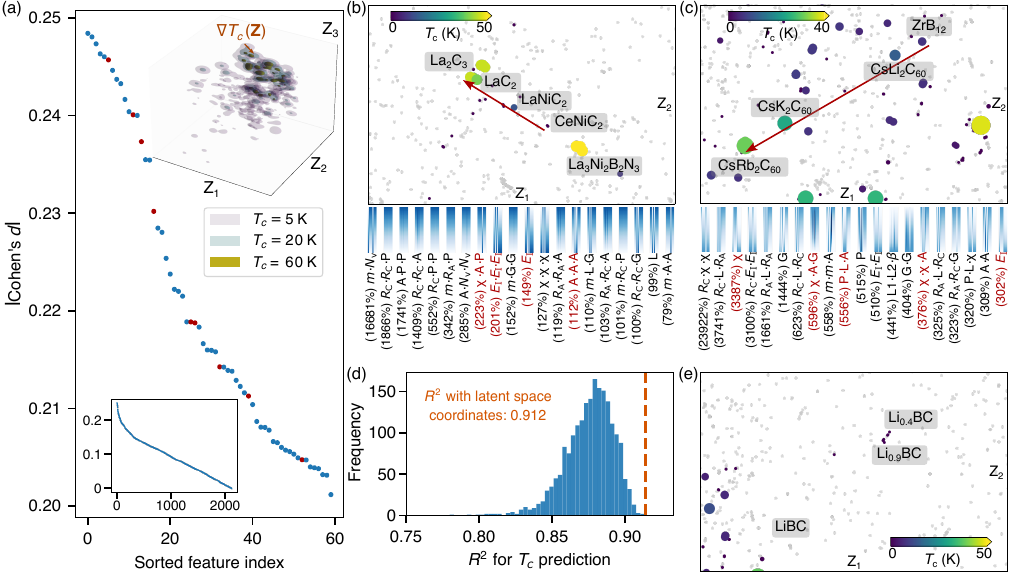}
\caption{{\bf Geometric interpretability of superconducting $T_c$}
(a)~Distribution of Cohen's $d$ values for the correlation between microscopic features and $
\nabla T_c$ (top inset) defined in latent space;
only $\approx$ 60 features have $d>0.2$ (bottom inset).
Local environment of superconducting clusters and trajectories of increasing $\nabla T_c$ obtained by slicing along the $Z_3$ axis and projecting onto the $Z_1$--$Z_2$ plane in (b) and (c).
The colored dots and their sizes represent superconducting $T_c$. The trajectories associated with the top $20$ decoded latent space features (based on Cohen's $d$ value; see red arrow) are highlighted in bottom panel.
(d)~Gaussian process prediction model for $T_c$  based on the three latent space features (orange), compared to models trained on 2{,}000 random combinations of three of the original features (blue), yields $R^2=0.912$.
(e)~Local environment of Li$_{x}$BC in the embedding, showing no proximity to high-$T_c$ superconductors.
%.
Our $\Gamma$AE prediction for their superconducting $T_c$ is close to the experimental values;
see text for details. 
}
\label{fig:interpretation}
\end{figure*}

To develop a quantitative connection between the features governing the variance across the materials in the low-dimensional embedding and superconducting $T_c$, we first define a smooth $T_c(\{Z_i\})$ ``field" in Fig.~\figref{fig:interpretation}{a}. Using a simple moving average or a radial basis function kernel yields similar results. We then compute $\nabla T_c$ everywhere in latent space and identify the directions associated with the largest variations in $T_c$, while correlating it with variations in the underlying microscopic features.  

To quantify these correlations, we evaluate the derivative of each feature along a unit vector pointing along $\nabla T_c$ and compare it against a randomly chosen direction that represents the baseline (i.e. ``null") value. For each feature $j$, we compare the following two distributions: $\{d_{j}^{\rm grad}\}$, which encodes variations along $\nabla T_c$, and $\{d_{j}^{\rm rand}\}$, encoding variations along the randomly chosen directions. We use the two-sample two-sided $t$-test~\cite{freund_statistical_2003} to evaluate the extent to which the two samples differ by accounting for both the means and variances of the distributions. The effect is quantified using Cohen's $d$~\cite{freund_statistical_2003}, a standardized measure of the difference between two means expressed in units of pooled standard deviation, where $d = 0.2$, $0.5$, and $0.8$ are conventionally interpreted as small, medium, and large effects, respectively.

The distribution of Cohen's $d$ values is shown in Fig.~\figref{fig:interpretation}{a}, where 2.9\% of the features have a modest effect ($d>0.2$). We have analyzed the top features correlated with $\nabla T_c$, sorted by their effect size (see Appendix~\ref{sec:gradient}). We exemplify the predictive power of these features by analyzing two trajectories with increasing $T_c$ in latent space, sliced along the $Z_3$ axis and projected onto the $Z_1$--$Z_2$ plane; see Fig.~\figref{fig:interpretation}{b--c}. Along each trajectory, we evaluate the evolution of the (decoded) original features and plot the top 20 features that vary the most (averaged over the 7 descriptors corresponding to each feature: first 5 moments and minimum and maximum values). Interestingly, we find that these predominant features overlap significantly with the set of 20 features with largest Cohen's $d$ [shown in red in Fig.~\figref{fig:interpretation}{b--c} and correspondingly in Fig.~\figref{fig:interpretation}{a}]. While some features show huge variations in these particular examples, we attribute this to specific materials chemistry; our emphasis here is on the {\it global} analysis that reveals the {\it universally} important features. Thus the latent space coordinates capture the salient features that determine the route to increasing $T_c$, even when derived using the coarse-grained field $T_c(\{Z_i\})$ and its gradient $\nabla T_c$ over the entire space.

Finally, we train a Gaussian process regression model~\cite{rasmussen_gaussian_2006} for predicting $T_c$ based on the three latent space coordinates. Our model achieves a test $R^2$ score of $0.912$, significantly high compared to a distribution of $2{,}000$ similar models trained on three randomly selected features from the original feature space; see Fig.~\figref{fig:interpretation}{d} (and also Fig.~\ref{fig:null_dist_2nd}).

Our unsupervised algorithm is able to learn the small subset of microscopic features that controls a material's superconducting character in an interpretable fashion. As we will now demonstrate, this enables a reliable and quantitatively accurate way of predicting $T_c$, without relying on the approximation schemes that arise in various first-principles-based or effective-model-based approaches. For instance, instead of using density functional theory (DFT)~\cite{kohn_self-consistent_1965,parr_density-functional_1994} and the approximate Migdal-Eliashberg theory~\cite{marsiglio_eliashberg_2020,sanna_combining_2020}, we can use our $T_c$ regression model based on \emph{only} three latent space features to predict $T_c$.

For example, if we focus on the family Li$_{x}$BC with $x\leq1$, our $T_c$ regression model based on the three latent space features predicts $T_c=1.5$--$8\,{\rm K}$, depending on the doping, which is consistent with experiments reporting $T_c\lesssim2$~K~\cite{bharathi_synthesis_2002,fogg_synthesis_2003}. The local environment of Li$_{x}$BC in our geometric embedding does not contain \emph{any} high-$T_c$ superconductors; see Fig.~\figref{fig:interpretation}{e}. This is especially interesting given the extra care required to predict $T_c$ from first principles in Li-based compounds~\cite{rosner_prediction_2002,fogg_chemical_2006,akashi_high-temperature_2012,borinaga_anharmonicity_2017,szewczyk_nonadiabatic_2020}.
This clearly demonstrates that if the crystal structures are available, our $\Gamma$AE model based on the low-dimensional geometric embedding provides a statistically reliable way of predicting $T_c$ in broad agreement with the experiment, without making any assumptions regarding the superconducting mechanism itself and based on an unsupervised generation of the geometric embedding.

\section{Discussion and Conclusion}\label{sec:discussion}
Theoretical understanding of collective phenomena in quantum materials typically relies on building effective electronic models and solving them using approximate methods. While successful for uncovering their universal aspects, such as conventional superconductivity, this approach leaves a fundamental gap in our understanding of materials design. The microscopic properties within the unit cell that govern these emergent phenomena are often unclear. Here we have presented a systematic theoretical framework for identifying the chemical and geometric features that collectively help differentiate between the macroscopic properties of electronic solids. Surprisingly, despite the complexity of unit cells containing tens of atoms, only a handful of covarying properties capture most of their variance.

Though we have primarily focused on superconductivity in this manuscript, our method generalizes to other phenomena. Future work can help unveil the features that determine the emergence of topological insulators, semimetals, magnets, and beyond, as long as a large enough dataset of experimentally verified properties is available~\cite{barroso-luque_open_2024,itani_northeast_2025,itani_large_2024,petralanda_two-dimensional_2024}. In the context of topological materials, this approach can be particularly promising, where conventional electronic structure computations often fail in the presence of strong correlations. Our unsupervised dimensionality reduction may help reveal the microscopic signatures of non-trivial topology obscured by traditional approaches.

The geometric interpretation of our algorithm enables rational materials design when combined with first-principles computation of the structure. By perturbing the neighborhood of clusters with desirable properties, one can verify the structural stability of candidate materials and confirm that optimized structures re-embed in the same cluster, thus providing a validated pipeline prior to experimental synthesis. Interestingly, the embeddings and associated minimal feature subset capturing the maximum variability can enable construction of effective low-energy models for the emergent properties of interest. An equally compelling direction is applying our framework to two-dimensional moir\'e materials~\cite{andrei_marvels_2021,nuckolls_microscopic_2024}, where unsupervised clustering of relaxed structures can bridge the knowledge gap between microscopic chemistry and structure on one hand, and emergent low-energy properties on the other, particularly given the rapid pace of experimental discoveries in the field.

\section*{Acknowledgments}
We thank T.~Arias, B.A.~Bernevig, B.~Ramshaw, I.~Cohen, and J.~Tahmassebpour for helpful discussions. O.L.\ and J.Z.K.\ acknowledge support from the Bethe-KIC postdoctoral fellowship at Cornell University. J.Z.K.\ acknowledges support from the Mong Neurotech postdoctoral fellowship and the Eric and Wendy Schmidt AI in Science Postdoctoral Fellowship at Cornell University. D.C.\ is funded in part by a NSF CAREER grant (DMR-2237522) and by a Sloan research fellowship.

%%%%%%%%%%%%%%%%%%%%%%%%%%%%%%%%%%%%%%%%%%%%
%%%%% APPENDIX %%%%%%%%%%%%%%%%%%%%%%%%%%%%%
%%%%%%%%%%%%%%%%%%%%%%%%%%%%%%%%%%%%%%%%%%%%

\appendix
\section{From a material to a feature vector}\label{sec:sm_featurization}
In this section, we provide some additional details on different schemes for turning the structural and electronic properties of a crystal into a feature vector. 
Our goal here is two-fold: to include as much information as possible, beyond just the chemical formula tied to the crystalline unit cell, while ensuring that unnecessary redundancies that do not capture any further variability in latent space are avoided. Ultimately, the dimensionality reduction algorithm should be able to filter out the collective coordinates that capture a significant variance in the data. 
The simplest material featurization is based on their chemical composition~\cite{kaplan_deep_2025}. Consider a 118-dimensional feature vector, $\vec{x}$, where an entry $x_j$ is the relative content of element $j$ (total number of stable elements$=118$). 
For example, in FeSe [Fig.~1(a) of the main text], $x_{26}=x_{34}=1$, since Fe and Se are element numbers $26$ and $34$, respectively. Similarly, in MgB$_{2}$, $x_{5}=2$, $x_{12}=1$, since B and Mg are element numbers $5$ and $12$, respectively. Apart from the elements in the chemical formula, all other $x_{j}=0$. Note that apart from the chemical composition, this featurization does {\it not} include information on the underlying structural or electronic properties within the unit cell. 

\begin{figure*}
    \centering
    \includegraphics[width=\linewidth]{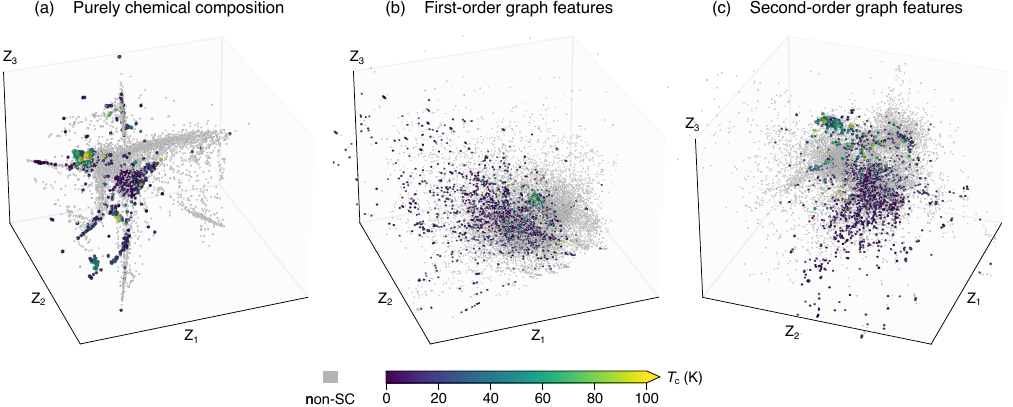}
    \caption{
    {\bf Clustering complexity across featurizations.} Three-dimensional embeddings for
    (a)~Purely compositional features (see Ref.~\cite{kaplan_deep_2025}).
    (b)~First-order graph-based features.
    (c)~First and second-order graph-based features.
    }
    \label{sfig:different_featurizations}
\end{figure*}

To incorporate more comprehensive information on a given compound, we take a graph-based approach, similar to Refs.~\cite{scarselli_graph_2009,isayev_universal_2017,bronstein_geometric_2021,reiser_graph_2022,fung_benchmarking_2021,choudhary_atomistic_2021,lesser_learning_2025,panigrahi_graphlet_2026}
Using the full information on the unit cell, as provided in a CIF file, we construct a graph based on the atomic locations and bonds in the unit cell to employ a hierarchical featurization of clusters. We choose $9$ chemical properties associated with individual atoms in the unit cell as our first-order features: atomic mass ($M$), atomic radius ($R_{\rm A}$), covalent radius ($R_{\rm C}$), Pauling electronegativity ($\chi$), electron affinity ($A$), first ionization energy ($E_{I}$), number of valence electrons ($N_{\rm v}$), group ($G$), and period ($P$) in the periodic table.
Since the unit cells of interest to us contain multiple atoms, we include the distribution of each atomic feature and calculate the first five (centralized) moments of their distribution, along with their minimal and maximal values.
At second order, we consider clusters of two atoms linked by a bond, and calculate the moments of the distributions of \emph{products} of atomic features, e.g., atomic mass of the first atom times covalent radius of the second atom. Furthermore, at second-order the bond length ($L$) also appears as a new \emph{geometric} feature, totaling 46 second-order features. Extending to third-order features leads to four types of distributions:
\begin{enumerate}
    \item Triple product of atomic features, e.g., atomic mass of the first atom times covalent radius of the second atom times electronegativity of the third atom (165 features).
    \item Two atomic features multiplied by the bond length connecting them, e.g., atomic mass of the first atom times covalent radius of the second atom times the length of the bond linking the two atoms (81 features).
    \item Bond angle ($\beta$), another geometric feature appearing for the first time in three-site clusters.
    \item Bond angle $\beta$ multiplied by the two bond lengths $L_1$, $L_2$ forming it.
\end{enumerate}
Overall, there are 248 third-order features.
For the three-dimensional embedding discussed in the main text, we include up to these third-order features. The resulting feature vector has $303\times(2+n_{\rm moments})$ entries, and since we take $n_{\rm moments}=5$, our feature vector has $2{,}121$ dimensions. In Appendix~\ref{sec:different_featurizations} below, we present results for the three-dimensional embedding based on only the first, and both first and second-order features, respectively. These are based on a feature vector with dimensions $9\times7=63$ and $55\times7=385$.

\section{A geometric low-dimensional manifold of materials}\label{sec:gvae}
To organize our understanding of the landscape of materials, two approaches are commonly taken. The first is to use high-dimensional featurization of materials datasets and perform feature selection for a particular material property~\cite{gashmard2024predicting}. The second is to perform unsupervised dimensionality reduction on subsets of materials to discover underlying organizing principles~\cite{park2025mapping}. However, nonlinear dimensionality reduction techniques such as UMAP~\cite{mcinnes2018umap}, tSNE~\cite{van2008visualizing}, and VAEs~\cite{kingma2019introduction} are known to significantly distort the geometry of the data due to the nonlinearities inherent in the methods, thereby leading to undesirable characteristics such as sensitivity of the embedding to initial conditions~\cite{kobak2021initialization} and distortions of the manifold geometry~\cite{meilua2024manifold}. To overcome these limitations and build a quantitative low-dimensional coordinate system for the space of materials, we leverage a recently developed manifold construction technique that uses differential geometry to control all of the nonlinearities involved in the manifold~\cite{kim2024gamma}.

\begin{figure*}
    \centering
    \includegraphics[width=\linewidth]{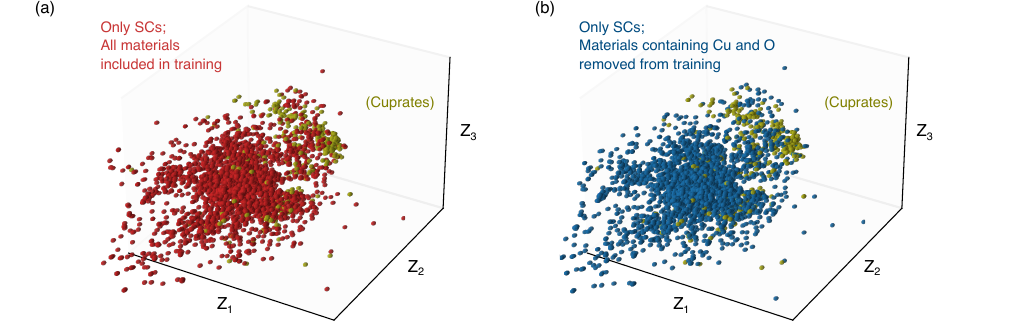}
    \caption{
    Robustness is demonstrated by comparing the embedding with all materials included in training (a) to an embedding trained without any materials that contain both Cu and O (b), which covers all cuprate superconductors. The two embeddings overlap almost perfectly (Pearson $r=0.991$ for pairwise distances; see Fig.~\ref{fig:re_embedding}).
    }
    \label{sfig:re_emb_cuo}
\end{figure*}

We regularize the geometry of the manifold by penalizing two curvatures: the \textit{parameter-effects curvature} that distorts the metric along the tangent bundle, and the \textit{extrinsic curvature} that allows the low-dimensional manifold to bend sharply, if the curvature is not accounted for, in new directions in materials space. 
To ameliorate these curvatures, we take the standard autoencoder loss function,
\begin{align}
    \mathcal{L}_{\mathrm{recon}}(\bm{x}) = \|\bm{x} - h_{\phi}(f_{\theta})(\bm{x}) \|_F^2,
\end{align}
and add a penalty for the extrinsic curvature,
\begin{equation}    
\begin{aligned}
    \mathcal{L}_{\mathrm{ext}} = \sum_a g^{\mu\bar{\mu}} & g^{\nu\bar{\nu}} \left(\frac{\partial^2\bm{h}_{\phi}^{a}}{\partial z^{\mu}z^{\nu}} - \Gamma_{\mu\nu}^{\kappa} \frac{\partial \bm{h}_{\phi}^{a}}{\partial z^{\kappa}}\right) \\ &\times\left(\frac{\partial^2\bm{h}_{\phi}^{a}}{\partial z^{\bar{\mu}}z^{\bar{\nu}}} - \Gamma_{\bar{\mu}\bar{\nu}}^{\bar{\kappa}} \frac{\partial \bm{h}_{\phi}^{a}}{\partial z^{\bar{\kappa}}}\right),
\end{aligned}
\end{equation}
which quantifies the bending of the manifold into new directions in materials space, and the parameter-effects curvature,
\begin{align}
    \mathcal{L}_{\mathrm{param}} = \sum_a g^{\mu\bar{\mu}}g^{\nu\bar{\nu}} \left(\Gamma_{\mu\nu}^{\kappa} \frac{\partial \bm{h}_{\phi}^{a}}{\partial z^{\kappa}}\right)
    \left(\Gamma_{\bar{\mu}\bar{\nu}}^{\bar{\kappa}} \frac{\partial \bm{h}_{\phi}^{a}}{\partial z^{\bar{\kappa}}}\right),
\end{align}
which quantifies the stretching of the manifold along the manifold tangent space. Here, $a$ indexes each dimension of the feature space, $z$ indexes the latent-space coordinates, $g^{\mu\nu}$ is the inverse of the metric tensor, and $\Gamma$ are the Christoffel symbols~\cite{choquet1982analysis}. Hence, our total loss function becomes
\begin{align}
    \mathcal{L} = \mathcal{L}_{\mathrm{recon}} + \alpha\mathcal{L}_{\mathrm{ext}} + \gamma\mathcal{L}_{\mathrm{param}},
\end{align}
where $\alpha$ and $\gamma$ are hyperparameters. Hence, at the limit of full complete regularization of both of these curvatures by making their cost go to zero, the manifold becomes flat, analogous to PCA. Our decomposition allows us to perturb the manifold geometry away from PCA to capture the emergent, nonlinear, low-dimensional landscape of materials in a quantitative and hypothesis-driven way. We also penalize related curvature-controlled losses for the encoder.

\section{Dependence on featurization schemes}\label{sec:different_featurizations}
In this section, we study the effects of increasingly complex featurization schemes on the resulting three-dimensional embeddings. As described in Appendix~\ref{sec:sm_featurization}, we begin by studying the embedding associated with the featurization based purely on chemical composition~\cite{kaplan_deep_2025}; see Fig.~\figref{sfig:different_featurizations}{a}. 
In Fig.~\figref{sfig:different_featurizations}{b} we show the embedding corresponding to the graph-based featurization truncated at first order (i.e. single atoms). Similarly, in Fig.~\figref{sfig:different_featurizations}{c} we show the embedding when the graph-based featurization is truncated at second order (i.e., containing both single atoms and two-atom clusters); see Appendix~\ref{sec:sm_featurization}. Across all the featurization schemes, we observe clustering of superconductors, with the composition-only featurization showing the most distinct geometric structure. 
Interestingly, with the introduction of graph-based features, the three clusters emerge at first order and become increasingly more segregated with more complexity.

\begin{figure*}
    \centering
    \includegraphics[width=\linewidth]{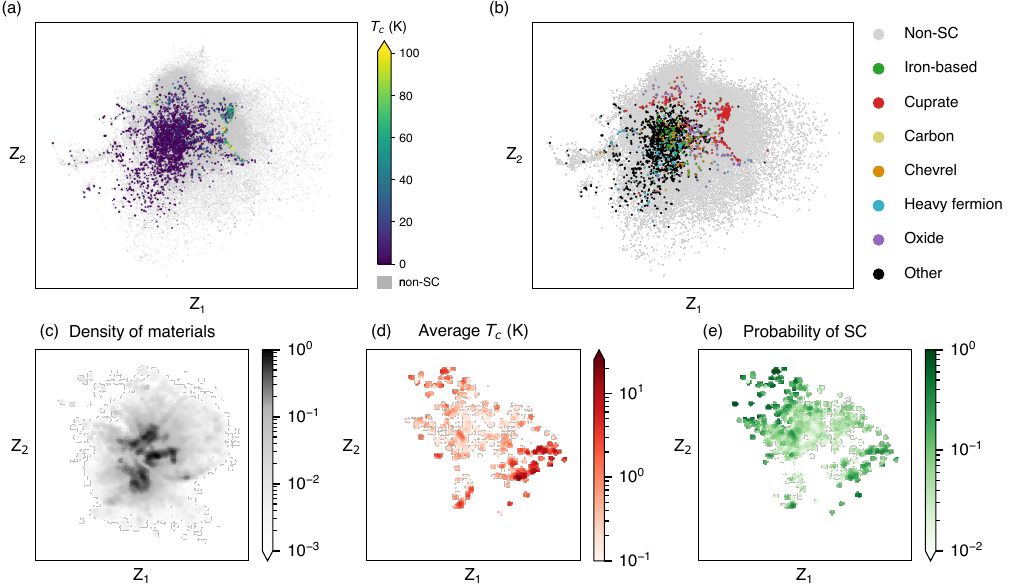}
    \caption{
    {\bf Two-dimensional embedding.} (a) The materials colored by their superconducting $T_c$, and (b) by their respective families.
    (c)~The density of materials across the embedding shows three emergent clusters.
    (d)~Average $T_c$, and (e) probability of finding a superconductor  across the embedding.
    The color scales are logarithmic for visualization purpose.
    }
    \label{sfig:2d}
\end{figure*}

We have performed the quantitative clustering analysis for all of the featurization schemes introduced here. Table~\ref{tab:zscore_embs} summarizes the results of the $z$-scored pairwise distances across each superconducting family, and for each embedding.
The $z$-score quantifies the number of standard deviations a data point is situated from the mean of the distribution, calculated as $z = (x - \mu)/\sigma$, where $x$ is the observed value, $\mu$ is the population mean, and $\sigma$ is the population standard deviation. For hypothesis testing, a sample mean's $z$-score is computed as $z = (\bar{x} - \mu)/(\sigma/\sqrt{n})$, where $\bar{x}$ is the sample mean and $n$ is the sample size. To determine if a sample is significantly smaller than the population distribution, a one-tailed test is performed where the null hypothesis is rejected if the calculated $z$-score falls below the critical value (e.g., $z < -1.645$ for $\alpha = 0.05$), indicating the sample mean is statistically significantly lower than the population mean.
We find that most SC families are significantly clustered for all schemes we have tried. The heavy fermion and other SC families are not clustered when using just first-order graph features, but when increasing the complexity to include also second- and third-order graph features, they become increasingly clustered.

We also extend the analysis shown in Fig.~\ref{fig:re_embedding} to further exemplify the robustness of the embedding. This time, we remove \emph{all} materials containing both Cu and O from the training; see Fig.~\ref{sfig:re_emb_cuo}. These are mostly layered materials, and they contain all cuprate superconductors. Notably, even with all these held out materials, the two embeddings overlap almost perfectly  (Pearson $r=0.991$ for pairwise distances). This demonstrates that $\Gamma$AE can capture the essence of cuprate superconductors without having seen anything resembling them.

\begin{table*}[b]
\centering
\begin{tabular}{lcccccc}
\hline
\textbf{Embedding} & \textbf{Iron-} & \textbf{Cuprate} & \textbf{Carbon} & \textbf{Chevrel} & \textbf{Heavy} & \textbf{Other} \\
 & \textbf{based} &  &  &  & \textbf{Fermion} & \textbf{SC}  \\
\hline
3D embedding, chemical composition features & $-10.01$ & $-43.41$ & $-11.05$ & $-1.38$ & $-15.88$ & $-48.77$ \\
2D embedding, third-order graph features & $-10.09$ & $-19.27$ & $-5.86$ & $-3.84$ & $-9.13$ & $-13.25$ \\
3D embedding, first-order graph features & $-16.16$ & $-27.91$ & $-4.24$ & $-7.75$ & $2.57$ & $11.62$ \\
3D embedding, second-order graph features & $-15.35$ & $-21.87$ & $-4.82$ & $-8.97$ & $-3.03$ & $-3.02$ \\
3D embedding, third-order graph features & $-11.63$ & $-20.66$ & $-5.64$ & $-5.00$ & $-7.02$ & $-11.36$ \\
\hline
\end{tabular}
\caption{Clustering of different families of superconductors in different embeddings, quantified by $z$-scored distances.}
\label{tab:zscore_embs}
\end{table*}

\section{Two-dimensional embedding}\label{sec:2d_emb}

To study the extent to which the results remain similar for varying latent-space dimensions, we now analyze a two-dimensional embedding. Recall that with the same GNN features, the two-dimensional embedding preserves 57\% of the variance in the data, compared to 34\% in PCA; both capture smaller variance compared to the corresponding three-dimensional embeddings (see Fig.~1(e) in the main text). 
The embedding is shown in Fig.~\ref{sfig:2d}, colored by $T_c$ and by the SC family. We observe a general clustering of SC families, similar to the three-dimensional embedding.
To further analyze the clusters quantitatively, we define a moving window of a fixed size to scan the latent space, and compute a variety of coarse-grained properties of interest. The simplest such quantity is the local density of materials within each window; see Fig.~\figref{sfig:2d}{c}. Interestingly, once again we find that the materials are largely organized into three high-density clusters. 

Focusing now on superconductivity, we evaluate average $T_c$ [Fig.~\figref{sfig:2d}{d}], and the probability of finding a SC [Fig.~\figref{sfig:2d}{e}] across the embedding. The latter is defined as the number of superconductors within the window divided by the total number of materials in that same window. While these two quantities approximately track each other, there are some subtle differences. For instance, there are clusters with high probability of superconductivity,  but with very low $T_c$ leading to a low average value. Interestingly, we do not find a clear positive correlation between these superconducting properties and the density of materials field; instead, the superconductors are largely restricted to areas with a relatively low density of materials.

\section{Gradient analysis}\label{sec:gradient}
Here we provide additional details on the computations that highlight the correlations between microscopic features and local gradients, $\nabla T_c$ in the three-dimensional embedding. This yields a set of the most predictive features, as evidenced by having the largest values of Cohen's $d$ when comparing against gradients along randomly chosen directions. Table~\ref{tab:features_cohens_d} lists the top 20 features according to this analysis.

Cohen's $d$ is a standardized effect size measure that quantifies the difference between two means in terms of pooled standard deviation, calculated as $d = (\bar{x}_1 - \bar{x}_2)/s_p$, where $\bar{x}_1$ and $\bar{x}_2$ are the sample means and $s_p$ is the pooled standard deviation. The pooled standard deviation is computed as $s_p = \sqrt{\frac{(n_1-1)s_1^2 + (n_2-1)s_2^2}{n_1 + n_2 - 2}}$ for unequal sample sizes, where $n_i$ and $s_i$ represent the sample sizes and standard deviations of each group. When comparing a sample to a population or reference distribution, Cohen's $d$ can be simplified to $d = (\bar{x} - \mu)/s$, with negative values indicating the sample mean is smaller than the reference mean. Conventionally, $0.2<|d|<0.5$ is interpreted as small to medium effect size.

\begin{table}[b]
\centering
\begin{tabular}{lcc}
\hline
\textbf{Feature} & \textbf{Type} & \textbf{Cohen's $d$} \\
\hline
$E_{\rm I}$$\cdot$$E_{\rm I}$$\cdot$A & Maximum & 0.249 \\
$E_{\rm I}$$\cdot$L$\cdot$$R_{\rm C}$ & 4th moment & 0.249 \\
$E_{\rm I}$$\cdot$L$\cdot$$R_{\rm A}$ & 4th moment & 0.248 \\
$\chi$$\cdot$$E_{\rm I}$$\cdot$A & Maximum & 0.247 \\
$\chi$$\cdot$$\chi$$\cdot$A & Maximum & 0.246 \\
$\chi$$\cdot$L$\cdot$$R_{\rm C}$ & 4th moment & 0.244 \\
$\chi$$\cdot$L$\cdot$$R_{\rm A}$ & 4th moment & 0.244 \\
$E_{\rm I}$ & Maximum & 0.240 \\
A & Maximum & 0.240 \\
$\chi$ & Maximum & 0.230 \\
A$\cdot$A$\cdot$A & 2nd moment & 0.225 \\
$N_{\rm v}$$\cdot$L$\cdot$$R_{\rm C}$ & 4th moment & 0.219 \\
$N_{\rm v}$$\cdot$L$\cdot$$R_{\rm A}$ & 4th moment & 0.217 \\
$E_{\rm I}$$\cdot$$E_{\rm I}$$\cdot$$N_{\rm v}$ & 1st moment & 0.214 \\
$\chi$$\cdot$L$\cdot$P & 2nd moment & 0.214 \\
$E_{\rm I}$$\cdot$L$\cdot$P & 2nd moment & 0.212 \\
$\chi$$\cdot$A$\cdot$P & 5th moment & 0.210 \\
$\chi$$\cdot$L$\cdot$A & 2nd moment & 0.208 \\
$E_{\rm I}$$\cdot$L$\cdot$A & Maximum & 0.208 \\
$E_{\rm I}$$\cdot$A$\cdot$A & Maximum & 0.208 \\
\hline
\end{tabular}
\caption{Top features according to Cohen's $d$ analysis when comparing their variation along the local gradient $\nabla T_c$ to their variation along randomly chosen directions.}
\label{tab:features_cohens_d}
\end{table}

\begin{figure}
    \centering
    \includegraphics[width=\linewidth]{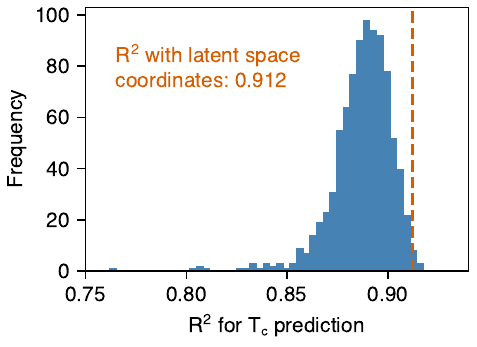}
    \caption{Gaussian process prediction model for $T_c$  based on the three latent space features (orange), compared to models trained on 900 random combinations of three of the original features (blue), using only second-order features.}
    \label{fig:null_dist_2nd}
\end{figure}

We have also examined the $R^2$ distribution for $T_c$ predictions, using only second-order features and no overlapping moments. This was done to ensure a non-dilute pool of descriptors. Figure~\ref{fig:null_dist_2nd} compares this distribution of performances in Gaussian process regression using random second-order features to the performance of regression based on the latent space features. We still find that the latent space features outperform the vast majority of combinations.

\bibliography{library}

\end{document}